\documentclass[pdflatex,sn-aps,iicol]{sn-jnl}% Basic Springer Nature Reference Style/Chemistry Reference Style

\usepackage{graphicx}%
\usepackage{multirow}%
\usepackage{amsmath,amssymb,amsfonts}%
\usepackage{amsthm}%
\usepackage{mathrsfs}%
\usepackage[title]{appendix}%
\usepackage{xcolor}%
\usepackage{textcomp}%
\usepackage{manyfoot}%
\usepackage{booktabs}%
\usepackage{algorithm}%
\usepackage{algorithmicx}%
\usepackage{algpseudocode}%
\usepackage{listings}%
\usepackage{graphicx}
\usepackage{amssymb,amsmath,verbatim,ulem}
	\usepackage{gensymb}
  \usepackage{nameref}
\usepackage{varioref}
\usepackage{hyperref}
\usepackage{cleveref}
\usepackage{natbib} %Bibliography.

\begin{document}

\title[Direct spectroscopy of Rubidium using a narrow-line transition at 420 nm]{Direct spectroscopy of Rubidium using a narrow-line transition at 420 nm}


\author[1]{\fnm{Rajnandan} \sur{Choudhury Das}}\email{d.rajnandan@iitg.ac.in}

\author[1]{\fnm{Samir} \sur{Khan}}\email{s.khan@iitg.ac.in}

\author[1]{\fnm{Thilagaraj} \sur{R}}\email{rthilagaraj@iitg.ac.in}
\author*[1]{\fnm{Kanhaiya} \sur{Pandey}}\email{kanhaiyapandey@iitg.ac.in}

\affil*[1]{\orgdiv{Department of Physics}, \orgname{Indian Institute of Technology Guwahati}, \orgaddress{\city{Guwahati}, \postcode{781039}, \state{Assam}, \country{India}}}


\abstract{The 5S$\to$6P transition in Rubidium (Rb) at 420 nm offers the advantage of a narrower linewidth and diverse applications in quantum technologies. However, the direct spectroscopy at this transition is challenging due to its weak transition strength. In this paper, we have discussed the saturated absorption spectroscopy (SAS) of Rb using the narrow-line transition at 420 nm. We have studied the effect of the temperature of the Rb cell, pump power and the beam size on the SAS dip heights and their linewidths. Additionally, our study offers a comprehensive examination, encompassing all eight error signals of Rb for the 5S$\to$6P transition at 420 nm and 421 nm. These findings contribute valuable insights to the field of laser frequency stabilization of Rb at blue transition and can be useful in quantum technologies based on this transition.}




\maketitle

\section{Introduction}\label{sec1}

In the intricate landscape of precision measurements and laser technologies, spectroscopy holds a pivotal role, particularly in the domain of laser frequency stabilization \cite{demtroder1982,metcalf99}. This precise control of laser frequencies is a key factor in diverse array of applications, ranging from understanding fundamental physics to the development of cutting-edge technologies such as atomic sensors, quantum computers, quantum simulators, and atomic clocks. Among the various elements explored in this pursuit, Rubidium (Rb) has been a focal point of investigation. The well-established 5S$\to$5P transitions at 780 nm and 795 nm have long dominated the spectroscopic landscape \cite{Khan2017,Wu2018JosaB}.
	
The 5S$\to$6P transition at the blue wavelength of 420 nm has received less attention, despite its manifold advantages, notably a narrower linewidth ($2\pi\times 1.4$ MHz) compared to conventional infrared transitions ($2\pi\times 6$ MHz). The unique attributes of this transition open avenues for lower-temperature magneto-optical traps \cite{rajnandan2023,Ding2022,rajnandan2024continuous}, paving the way for applications in laser cooling and Rydberg excitation of Rb atoms \cite{Morsch2018,Faoro2016PRA,Valado2016PRA}. Its versatility extends further into the realms of single atom array \cite{Levine2018PRL,Ebadi2021}, quantum gate implementations \cite{Bluvstein2022}, improved atom interferometer \cite{Salvi2023PRL}, compact Rb optical frequency standards \cite{Zhang2017}, creation of narrow-bandwidth Faraday optical filters \cite{Ling2014OL,Guan2023IEEE}, demonstration of non-linear magneto-optical rotation \cite{Budker2015} and study of various atomic systems \cite{Boon1998,Ogaro2020JphyB,Ogaro2021PRA,Shylla2022EPJD}. Despite its potential, comprehensive understanding of the Rubidium spectrum at the blue transition has been surprisingly limited.

Traditionally, stabilizing the laser frequency at 420 nm involves double resonance spectroscopy \cite{rajnandan2023} or saturated absorption spectroscopy (SAS) \cite{Ding2022}. In double resonance spectroscopy, a V system is formed with a 780 nm probe and 420 nm control and electromagentically induced transperency (EIT) and optical pumping methods are used \cite{Ogaro2020JphyB}. However, this approach comes with complexities. The stability of the 420 nm signal depends on the stability of the 780 nm laser frequency, making it prone to noise and destabilization. Changes in the 780 nm laser frequency impact the 420 nm spectrum, adding complexities to experimental setups. Additionally, its absorption dip has large linewidth (typically $> 6$ MHz) due to the Doppler shift mismatch between the 780 nm and 420 nm lasers \cite{Ogaro2020JphyB,Ogaro2021PRA}. Moreover, it is highly sensitive to the angle between the 780 nm and 420 nm laser beams \cite{Ogaro2021PRA}.

%On the otherhand, SAS does not have these challenges and is widely used in laser spectroscopy \cite{demtroder1982,das2006precise,banerjee2003saturated,das2006Epjd}. SAS at narrow transition is challenging due to the weak transition strength as compared to the broad transitions. The vapor cell temperature need to be increased to $50-100~ ^0$C to improve the signal-to-noise ratio of the spectrum. This can result in coating on the vapor cell window, demanding careful consideration in experimental design and execution. SAS of Cs at narrow transition is widely studied \cite{Gerhardt1978,Wang2013Cs,Miao2022Cs,,Zhang2016Cs1,Zhang2016Cs2}. SAS of Rb at 420 nm requires blue color sensitive photodetectors which are now widely accessible. Various groups have also contributed to the partial study of the 5S$\to$6P spectrum of Rb\cite{Hemmerich1990OL, hayasaka2002, Chang2019}. Glaser et al. have conducted precise measurements of absolute transition frequencies at 420 nm and 421 nm, reporting all the eight Rb spectra \cite{Glaser2020}. The SAS dips exhibited typical linewidths of 2.7-2.8 MHz. A detailed investigation on the behavior of the Rb spectra at blue transition with the experimental parameters are still missing. In this work, we delve into the the effect of temperature of the vapor cell, pump power, and beam size on the SAS signal, providing a comprehensive analysis. Additionally, we present all eight error signals of Rb, corresponding to the 5S$\to$6P transition at 420 nm and 421 nm.

On the other hand, SAS avoids the challenges linked with double resonance spectroscopy. While SAS is a widely employed technique \cite{demtroder1982,das2006precise,banerjee2003saturated,das2006Epjd}, applying it to narrow transitions poses challenges due to the weaker transition strengths compared to broader transitions. Elevating the vapor cell temperature to $50-100~^\circ$C enhances the signal-to-noise ratio but introduces the possibility of coating on the vapor cell window, necessitating meticulous considerations in experimental design and execution. SAS on Cs at narrow transitions has been extensively studied \cite{Gerhardt1978, Wang2013Cs,Miao2022Cs, Zhang2016Cs1, Zhang2016Cs2}. For SAS on Rb at 420 nm, the use of blue color-sensitive photodetectors is essential which is now widely accessible. Glaser et al. have conducted precise measurements of absolute transition frequencies at 420 nm and 421 nm, reporting all the eight Rb spectra \cite{Glaser2020}. The SAS dips exhibited typical linewidths of 2.7-2.8 MHz. While various groups have made partial contributions to the study of the 5S$\to$6P spectrum of Rb \cite{Hemmerich1990OL, hayasaka2002, Chang2019}, a comprehensive investigation into the behavior of Rb spectra at the blue transition concerning experimental parameters is still lacking. This work addresses this gap by exploring the effects of temperature of the vapor cell, pump power, and beam size on the SAS signal. Additionally, we present all eight error signals of Rb corresponding to the 5S$\to$6P transition at 420 nm and 421 nm, providing a thorough analysis of this atomic transition.

\section{Experimental Set-up}\label{ExpSetUp}
\begin{figure}
\centering
\includegraphics[width=1\linewidth]{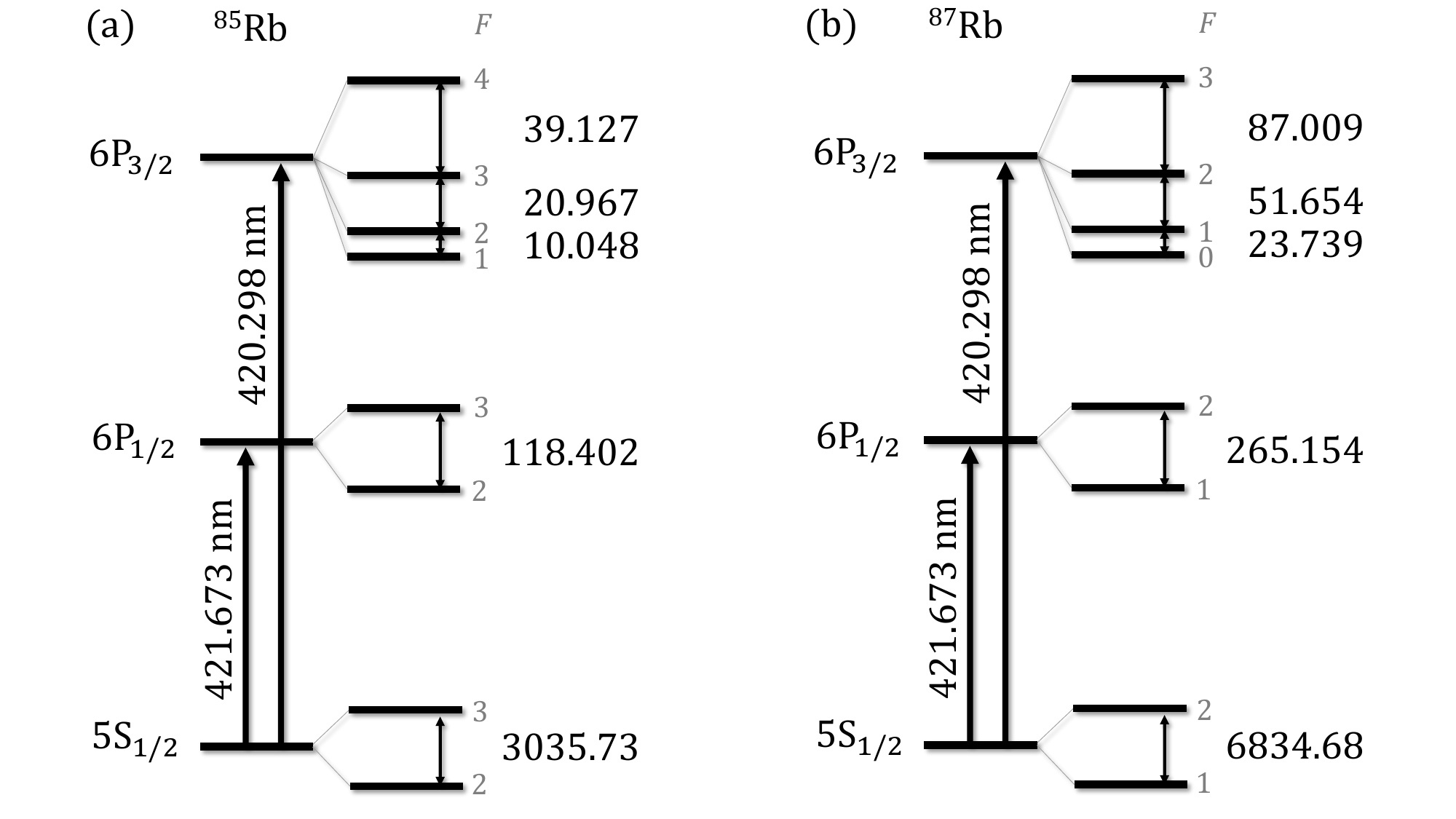}
\caption{The relevant energy level diagram and hyperfine splitting (in MHz) for the 5S$_{1/2}$ and 6P states of (a) $^{85}$Rb and (b) $^{87}$Rb.}
\label{BlueEnergyLevels} %used for image reference
\end{figure}

The experimental set-up comprises one commercially available external cavity diode laser (ECDL) from Toptica Photonics with model no. DL pro HP. It has a coarse tuning range of 420 - 423 nm and a typical linewidth of $<200$ kHz. It has an integrated optical isolator to protect the laser diode from the back reflections from the other optical elements. The total available output power is 70 mW, and the output beam diameter is 3 mm $\times$ 4 mm. In the first half of the experiment, the wavelength is tuned to $420.298$ nm to observe the D$_2$ line of the 5S$_{1/2}\rightarrow$ 6P$_{3/2}$ Rb spectrum and in the next half, it is tuned to $421.673$ nm to observe the D$_1$ line of the 5S$_{1/2}\rightarrow$ 6P$_{1/2}$ spectrum. The relevant energy level diagram and hyperfine splitting (in MHz) for the 5S$_{1/2}$ and 6P states of $^{85}$Rb and $^{87}$Rb are shown in Fig. \ref{BlueEnergyLevels}. A leak beam from the blue laser is sent to a wavelength meter (make: Highfinesse GmbH, model: WS7-60) to monitor the single-mode operation and the wavelength of the blue laser.

\begin{figure}
\centering
\includegraphics[width=1\linewidth]{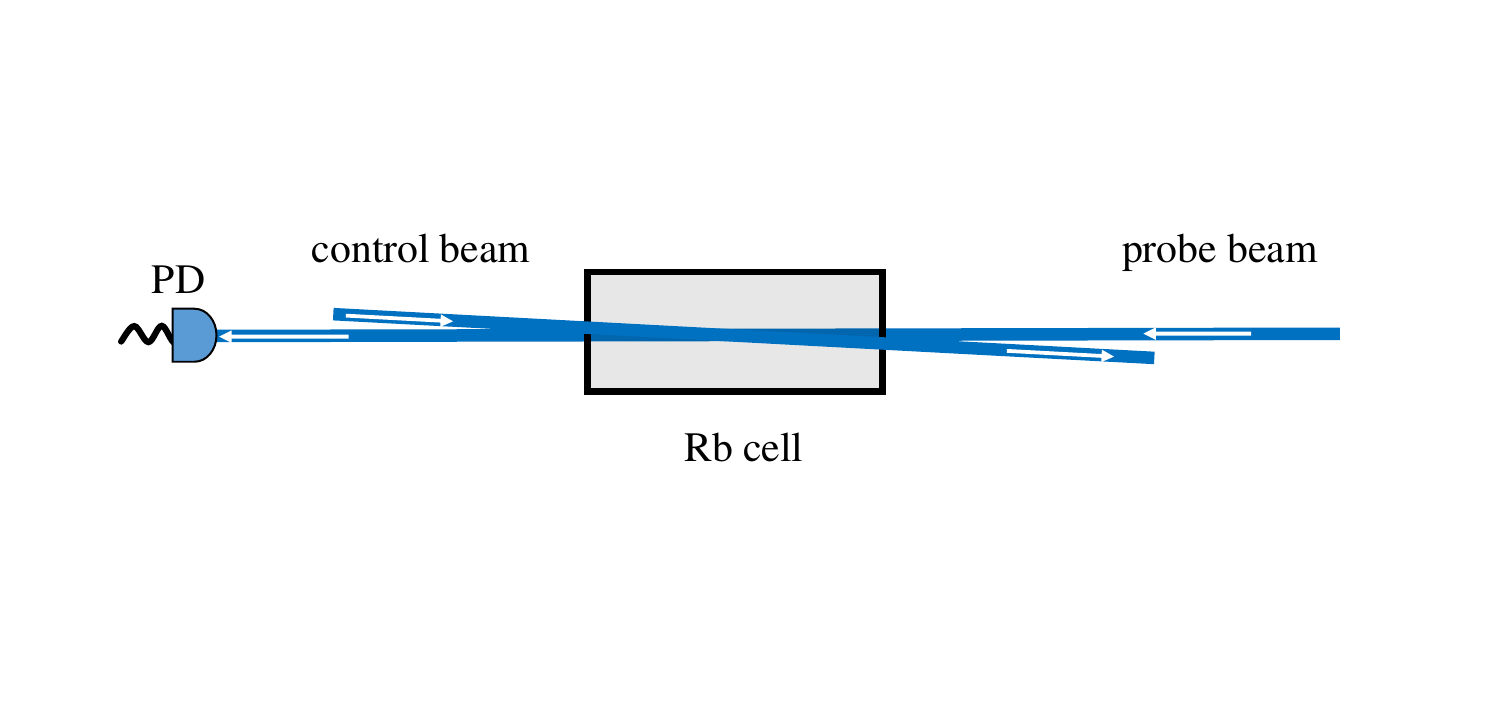}
\begin{picture}(0,0)
    \put(-100,110){(a)}
\end{picture}
  \includegraphics[width=1\linewidth]{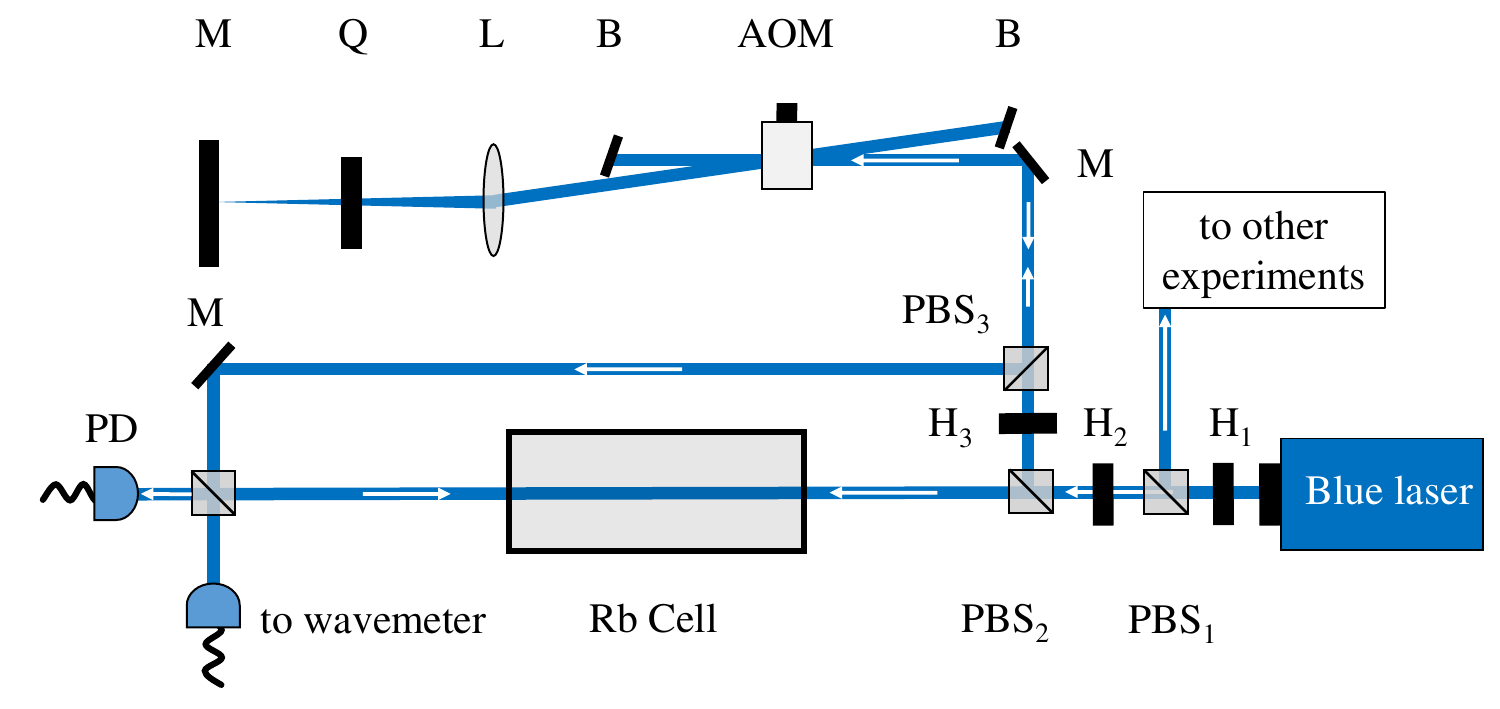}
  \begin{picture}(0,0)
    \put(-100,130){(b)}
\end{picture}
  \caption{\label{fig:ExptSetUpBlue}(Color online) Saturated absorption spectroscopy set-up for the 420 nm laser. (a) Basic set-up, (b) Detailed. Figure abbreviations: AOM: acousto-optical modulator, B: beam blocker, H$_1$, H$_2$, H$_3$: $\lambda/2$ wave-plates, L: lens, M: mirrror, PBS$_1$, PBS$_2$, PBS$_3$: polarizing beam splitter, PD: photo-detector, Q: $\lambda/4$ wave-plate. }
\end{figure}
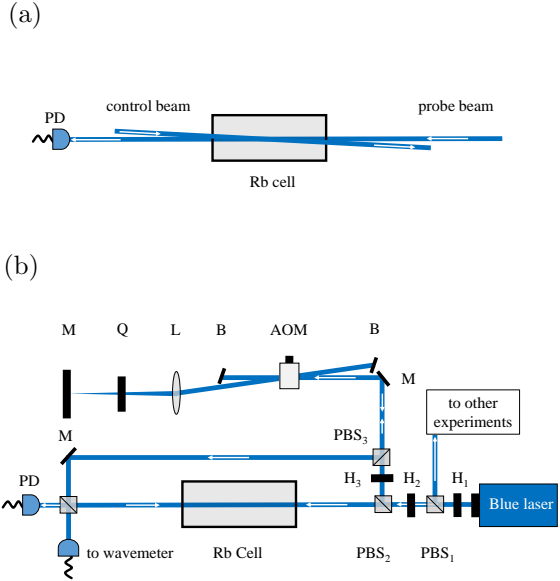

Fig. \ref{fig:ExptSetUpBlue}(a) depicts the schematics of the experimental set-up for the saturated absorption spectroscopy (SAS) of the $^{87}$Rb atom at blue transition. It consists of a rubidium vapor cell of length 100 mm and diameter 25 mm. The cell is wrapped with a thick layer of aluminium foil and placed inside a 150 mm long hollow cylindrical oven made of brass with an inner diameter of 27 mm and an outer diameter of 40 mm. Two hollow circular brass plates of the same inner and outer diameter are attached to both ends of the oven. Two quartz plates are attached on both ends of the oven to block the circular aperture and thermally isolate the vapor cell from the environment to avoid coating on the windows without limiting the optical access. The oven is wrapped with heating tape and another thick layer of aluminium foils. Current is passed through the heating tape using a variac to increase the oven's temperature. The temperature of the vapor cell is monitored using three thermocouples attached to its surface.

The blue laser is divided into two parts using the $\lambda/2$ wave-plate (H$_1$) and polarizing beam splitter (PBS$_1$), as shown in Fig. \ref{fig:ExptSetUpBlue}(b). The reflected beam from the PBS$_1$ is used for other experiments (not used in this work). The transmitted beam is used for the SAS and is divided into two beams using H$_2$ and PBS$_2$. The transmitted beam from PBS$_2$ is used as probe beam. It is sent through the Rb vapor cell and is detected on a UV-enhanced Si Variable-Gain Avalanche photo-detector of bandwidth DC - 400 MHz (make: Thorlabs, model: APD430A2/M). The reflected beam from PBS$_2$ is sent through H$_3$ and PBS$_3$.  The transmitted beam from PBS$_3$ is upshifted by $2\times46.75$ MHz using an AOM in a double pass configuration. Polarization of the beam is rotated by 90$^0$ by the $\lambda/4$ wave-plate (Q) and thus gets reflected beam from PBS$_3$. It is used as a control beam. It is mixed with the probe beam on the PBS placed in front of the photo-detector and is sent to the vapor cell with counter-propagating to the probe beam. The polarization of the probe and control beam are orthogonal to each other.  The power of the control beam is varied using H$_3$ and the PBS.

The laser frequency is tuned near the resonance and is scanned using piezo to find the SAS signal. It is observed and recorded on a digital storage oscilloscope of bandwidth 100 MHz. While the basic experimental schematics, as shown in Fig. \ref{fig:ExptSetUpBlue}(a), are sufficient for the detailed investigation of the behavior of the Rb spectra at the blue transition with experimental parameters, the additional AOM on the control laser beam path, as depicted in Fig. \ref{fig:ExptSetUpBlue}(b), is necessary to generate the error signal for frequency stabilization of the blue laser. To get the error signal, the rf frequency to the AOM is modulated at 10 kHz. The modulation is turned off, while studying the SAS dip height and linewidth with different parameters.

\section{Results and Discussion}{\label{RnD}}

\begin{figure}[t!]
    \centering
      \includegraphics[width=0.8\linewidth]{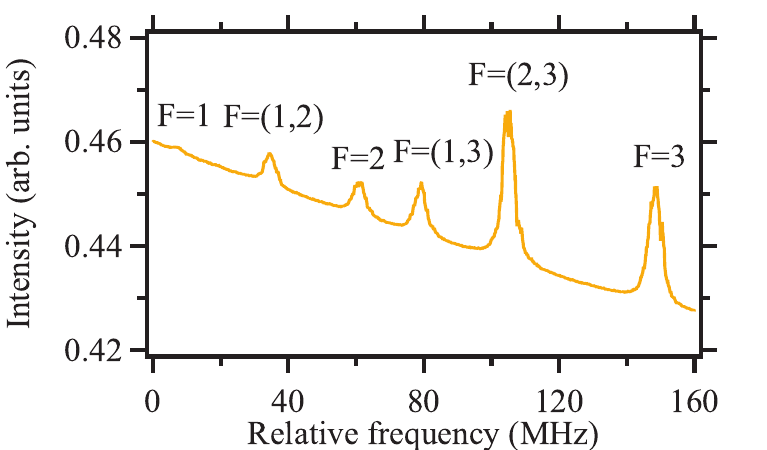}
      \begin{picture}(0,0)
        \put(-190,100){(a)}
      \end{picture}
              \includegraphics[width=0.8\linewidth]{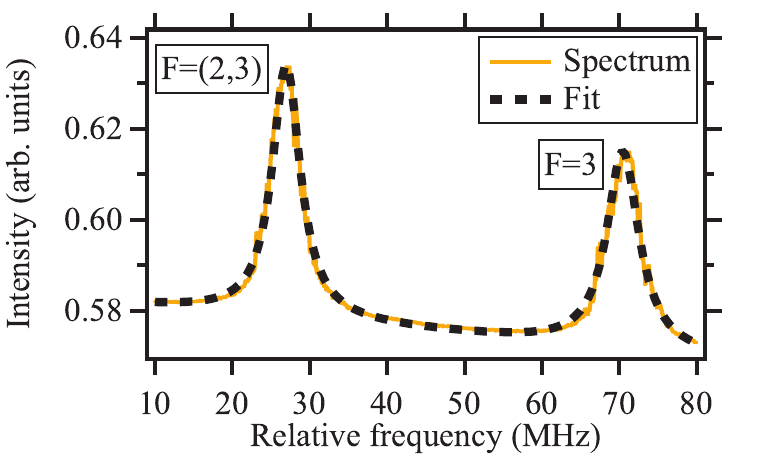}
              \begin{picture}(0,0)
                \put(-190,100){(b)}
              \end{picture}
      \caption{\label{Spectrum}(Color online) Saturated absorption spectroscopy spectrum for 5S$_{1/2}$(F$=$2)$ \rightarrow$ 6P$_{3/2}$ transition of $^{87}$Rb at 420 nm. (a) Full spectrum. The peaks correspond to F=1, (1,2), (2), (1,3), (2,3) and 3, from left to right. (b) Spectrum with $F=(2,3)$ cross-over and $F=3$ peaks. The orange line represents the spectrum and the black dotted line represents the fit.}
    \end{figure}
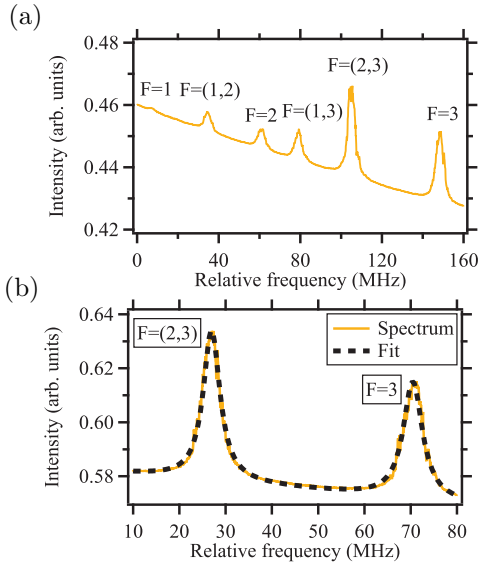

In the absence of the control beam, atoms get excited by absorbing the probe beam, resulting in a Doppler-broadened Spectra. In the presence of the control beam, the atoms get excited, absorbing the control beam. Thus, there is a reduction in the probe absorption, resulting in several dips in Doppler-broadened spectra. These dips correspond to the hyperfine transitions of the Rb atoms and their cross-overs. 

Initially, the blue laser is tuned to the 5S$_{1/2}$, F$=2 \rightarrow$ 6P$_{3/2}$ transition of $^{87}$Rb at 420 nm. It is scanned using piezo to obtain all three real peaks and three crossover peaks. A typical SAS signal at 80 $^0$C is shown in Fig. \ref{Spectrum}(a). 6 peaks corresponding to the transition from the lower ground state of $^{87}$Rb i.e., 5S$_{1/2}$ (F $=2$) to the excited state, 6P$_{3/2}$ (F $=X$) are identified within the Doppler envelope, where $X=$ 1, (1,2), 2, (1,3), (2,3) and 3 from left to right. Here, F $=(m, n)$ peak corresponds to the crossover resonance peak resulting from the F $=m$ and F $=n$ peaks. These crossovers are because when laser frequency is equal to the sum of half of the two excited state frequencies, both the control and probe beam interact with the same group of atoms, resulting in a dip halfway between two real Lamb dips. The x-axis is scaled in such a way that the separation between the F$=(2,3) $ and F$=3$ peaks becomes $43.5$ MHz \cite{Glaser2020}. 

\begin{figure}[t!]
    \centering
    \includegraphics[width=0.8\linewidth]{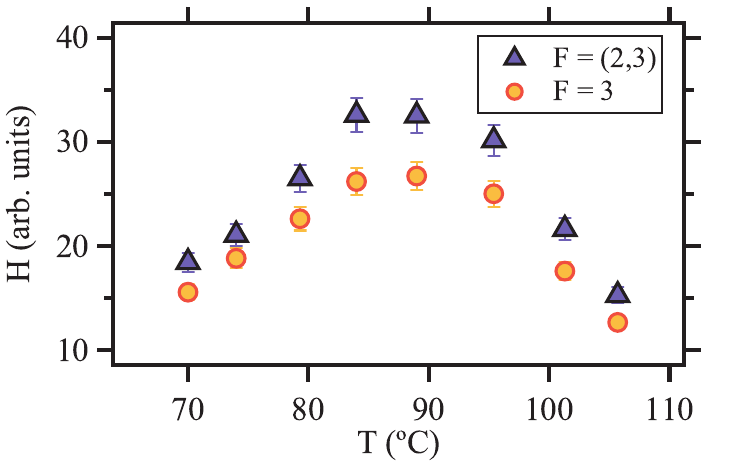}
      \caption{\label{Temp}(Color online) SAS dip height (H) vs the temperature of the Rb vapour cell (T). Purple triangle and orange circle corresponds to the $F=(2,3)$ cross-over and $F=3$ peaks, respectively.}
    \end{figure}

    \begin{figure}[t!]
        \centering
              \includegraphics[height=3.5cm]{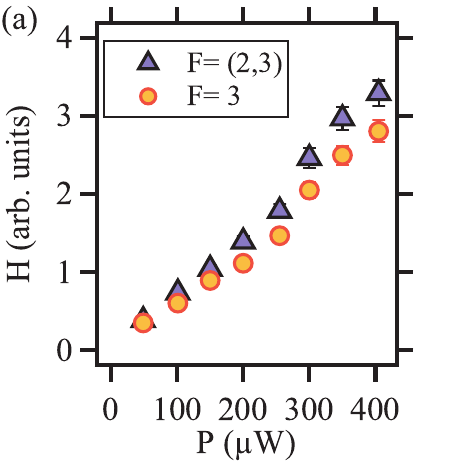}
        \includegraphics[height=3.5cm]{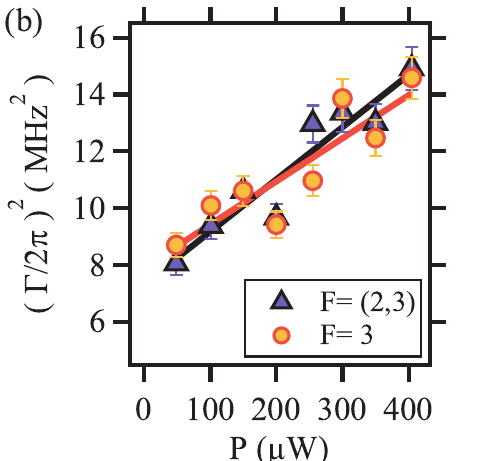}
          \caption{\label{Power}(Color online) (a) SAS dip height (H), and (b) square of the linewidth ($[\Gamma/2\pi]^{2}$) of the corresponding peaks vs the power of the control laser ($P$). Purple triangle and orange circle corresponds to the $F=(2,3)$ cross-over and $F=3$ peaks, respectively.}
        \end{figure}
    
Next, we focus on the two relatively larger peaks, i.e. F $=(2,3)$ and F $=3$. We decrease the scan amplitude to zoom into these two peaks. Typical SAS signal at 80 $^0$C are shown in Fig. \ref{Spectrum}(b) (orange line). The x-axis is scaled as discussed above. To measure the SAS dip height and the linewidth of the F$=$(2,3) and F$=$3, the spectrum is fitted with the equation $I=y_0+m\nu+ {A_{1}}/{[1+(\frac{2(\nu-\nu_0)}{\Gamma_1})^2]}+{A_{2}}/{[1+(\frac{2(\nu-\nu_0-43.5)}{\Gamma_2})^2]}$. Here, $I$ is the probe absorption, $\nu$ is the frequency of the 420 nm laser, $\nu_0$ is the location of F$=$(2,3) peak, $y_0+m\nu$ is the linear Doppler profile over which the F$=$(2,3) and the F$=3$ peaks are appearing, $A_{1(2)}$ and $\Gamma_{1(2)}$ are the peak height and linewidth of the F$=$(2,3) ($=$3) peak respectively. The black dashed line in \ref{Spectrum}(b) represents the fit. The linewidth of the F$=$(2,3) and the F$=3$ peaks are 4.57(3) MHz and 4.46(4) MHz respectively.
    
\begin{figure*}[t]
    \centering	
    \includegraphics[width=\linewidth]{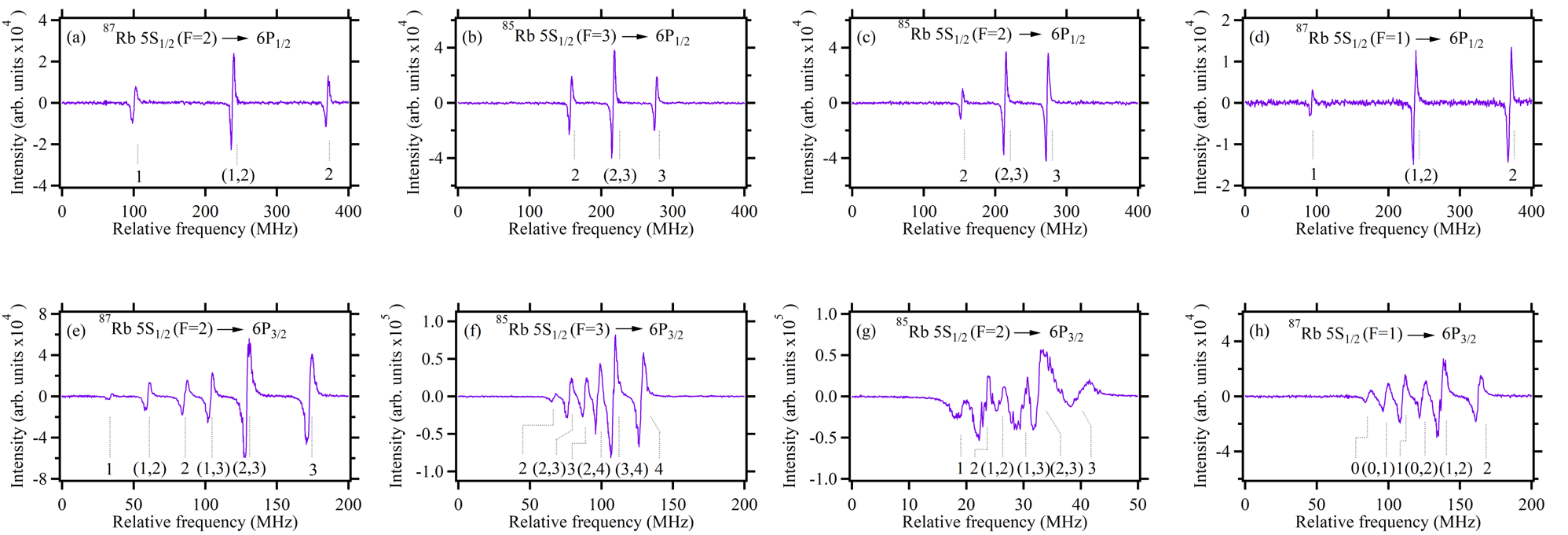}
    \caption{Recorded error signal corresponding to the 5S$_{1/2}\rightarrow$ 6P$_{1/2(3/2)}$ transition at 421 (420) nm.}
    \label{6p32}	
    \end{figure*}

We first study the effect of the temperature of the Rb vapor cell ($T$) on the SAS dip height ($H$) of the F$=$(2,3) and the F$=3$ peaks (shown in Fig. \ref{Temp}). The power of the probe and pump beams are 80 $\mu$W and 380 $\mu$W, respectively, and kept constant throughout the measurement. At room temperature, the spectrum could not be observed due to the weak transition strength of the blue transition. The temperature of the Rb vapor cell is slowly increased. We observe that the probe absorption increases as the temperature of the vapor cell is increased from 70 $^0$C to 84 $^0$C due to the increase in the number of atoms interacting with the beam inside the vapor cell. It then reaches a maximum and decreases as the temperature is further increased from 84 $^0$C to 106 $^0$C due to increased collisions between the atoms. Throughout the measurement, the height of the crossover peak, F $= (2,3)$ is larger than that of the F $=3$ peak. In the following portion of the experiment, the Rb vapor cell temperature is kept at 84 $^0$C since this results in the best signal-to-noise ratio.

%Linewidth of the spectrum initially increases from 3.6 MHz to 4 MHz in the range [70 $^0$C - 84 $^0$C] and then decreases to 3.2 MHz as the temperature is increased further. The initial increase in linewidth with temperature is due to the collisional broadening effect. Further decrease in linewidth with temperature is subject to future study(??????). It is observed that there is a competition between the height of the probe absorption and its linewidth. Although the linewidth is minimum (i.e. 3.2 MHz) at 106 $^0$C, the height of the probe absorption also minimum. 

Next, we investigate the effect of the power of control beam ($P$) on the SAS dip height ($H$) and linewidth ($\Gamma$) of the F$=$(2,3) and the F$=3$ peaks (shown in Fig. \ref{Power} (a) and (b)). The probe power is kept constant at 80 $\mu$W. We observe that with increase in power of pump beam, SAS dip height and linewidth increases due to increase in absorption and power broadening respectively. Due to power broadening mechanism, the linewidth ($\Gamma$) increases with increasing power as per the equation: $\Gamma = \Gamma_0\sqrt{1+I/I_s}$, where $\Gamma_0$ is a combination of natural broadening and the collisioanal broadeing, I is the intensity of the control beam and $I_s$ is the saturation intensity. We have done the least square fitting of the (linewidth)$^2$ vs power data as shown in Fig. \ref{Power} (b). By extrapollating to zero control power, we found that the minimum linewidth of F $= (2,3)$ and F $= 3$ peaks are 2.7 MHz and 2.8 MHz respectively. This is around 2 times the natural linewidth of the 6P$_{3/2}$ state (natural linewidth = 1.4 MHz). The difference in experimental measurement of the linewidth and the natural linewidth due to residual broadening mechanisms.

Next we study the effect of beam size on the peak height and its linewidth. We increase the size of the beams by 2 times with the help of combination of two plano-convex lenses of f=25 mm and f=50 mm. We increase the power of the pump and probe beam by 4 times to keep the same ratio of the intensities of the pump and probe beam. We observe simillar behaviour (as in Fig. \ref{Power}). However for the same pump to probe beam intensity ratio, the peak height increases by 2 times with the increased beam size. This is due to increase in number of atoms interacting with the blue beam. We also observe that the minimum linewidth of the F= (2,3) and F= 3 peaks are 4.4 MHz and 4.2 MHz, which is around 1.5 times more than the linewidth measured with 2x smaller beam size. %This indicates that the other broadening mechanisms such as collisional boadening mechansim is playing a role, due to which the linewidth measured is not close to the natural linewidth.

We extend this study to obtain the error signals by modulating the rf frequency of the AOM at 10 kHz. The error signal corresponding to the 5S$_{1/2}\rightarrow$ 6P$_{1/2}$ transition at 421 nm are shown in the first row of Fig. \ref{6p32}. (a) and (d) refers to the error signal of $^{87}$Rb corresponding to the upper ground state (F=2) and lower ground state (F=1), respectively. (b) and (c) refers to the error signal of $^{85}$Rb corresponding to the upper ground state (F=3) and lower ground state (F=2), respectively. The error signal corresponding to the 5S$_{1/2}\rightarrow$ 6P$_{3/2}$ transition at 420 nm are shown in the second row of Fig. \ref{6p32}. (a) and (d) refers to the error signal of $^{87}$Rb corresponding to the upper ground state (F=2) and lower ground state (F=1), respectively. (b) and (c) refers to the error signal of $^{85}$Rb corresponding to the upper ground state (F=3) and lower ground state (F=2), respectively. The errors signals in (a), (b) and (d) are well separated while in (c), they are slightly distorted due to closely spaced energy levels ($31$ MHz spread). All the errors signals are shown in increasing order of frequency.

Power of the probe beam and control beam and the allignments are kept unchanged while recording all the eight error signals. Thus, a rough comparison of the transition strength of the different peaks can be done. In contrast to the work by Glaser et. al \cite{Glaser2020}, we observe that in Fig. \ref{6p32}(a), (b) and (d), the amplitude corresponding to upper excited state is noticably larger than that corresponding to the lower excited state. We find the ratios of the amplitude of the the uper excited state to the lower excited state in (a)-(d), which are 1.4, 0.9, 3.8 and 4.4 respectively. These values are roughly equal to the theoretical values of the transition strength of the corresponding transitions (1, 0.8, 3.5 and 5 respectively) \cite{metcalf99}.

\section{Conclusions}{\label{Conclusion}}

In summary, our study delves into saturated absorption spectroscopy (SAS) of Rubidium using a narrow-line transition at 420 nm. We systematically examined the impact of temperature of the Rb cell, pump power, and beam size on SAS dip heights and linewidths. Achieving optimal signal-to-noise ratio at a Rb cell temperature of $84~\degree$C, we obtained minimum linewidths of 2.7 MHz and 2.8 MHz for the 5S$_{1/2}$, F=2$\rightarrow$ 6P$_{3/2}$ transitions with F= (2,3) and F= 3, respectively. Notably, doubling the beam size could enhance the signal-to-noise ratio twofold, albeit with a 1.5 times increase in spectrum linewidth. Furthermore, we presented all eight error signals for $^{85}$Rb and $^{87}$Rb corresponding to the 5S$_{1/2}$, F=2$\rightarrow$ 6P$_{3/2}$ transition at 420 nm and the 5S$_{1/2}$, F=2$\rightarrow$ 6P$_{1/2}$ transition at 421 nm. These findings significantly contribute to the foundational knowledge of SAS at the 420 nm transition in Rb, advancing laser frequency stabilization and is used for producing laser cooled Rb atoms at blue transition \cite{rajnandan2024continuous}. The insights gained from this study hold practical implications for applications in quantum technologies, particularly those based on the blue atomic transition of Rb.

\backmatter

\bmhead{Acknowledgements}

RCD would like to acknowledge the Ministry of Education, Government of India, for the Prime Minister's Research Fellowship (PMRF). K.P. would like to acknowledge the funding from DST through Grant No. DST/ICPS/QuST/Theme-3/2019.

%\setcitestyle{numbers} %Cite as numbers or author-year.
%\bibliographystyle{vancouver} %Reference style.
\bibliography{sn-bibliography}% common bib file

\begin{thebibliography}{10}
\providecommand{\url}[1]{{#1}}
\providecommand{\urlprefix}{URL }
\providecommand{\doi}[1]{\url{https://doi.org/#1}}
\bibcommenthead

\bibitem{demtroder1982}
W.~Demtr{\"o}der, \emph{Laser spectroscopy}, vol.~2 (Springer, 1982)

\bibitem{metcalf99}
H.J. Metcalf, P.~van~der Straten, \emph{Laser Cooling and Trapping} (Springer
  New York, NY, New York, 1999).
\newblock \urlprefix\url{https://doi.org/10.1007/978-1-4612-1470-0}

\bibitem{Khan2017}
S.~Khan, M.P. Kumar, V.~Bharti, V.~Natarajan, Coherent population trapping
  (cpt) versus electromagnetically induced transparency (eit).
\newblock The European Physical Journal D \textbf{71}(2), 38 (2017).
\newblock \doi{10.1140/epjd/e2017-70676-x}.
\newblock \urlprefix\url{https://doi.org/10.1140/epjd/e2017-70676-x}

\bibitem{Wu2018JosaB}
B.~Wu, Y.~Zhou, K.~Weng, D.~Zhu, Z.~Fu, B.~Cheng, X.~Wang, Q.~Lin, Modulation
  transfer spectroscopy for d1 transition line of rubidium.
\newblock J. Opt. Soc. Am. B \textbf{35}(11), 2705--2710 (2018).
\newblock \doi{10.1364/JOSAB.35.002705}.
\newblock
  \urlprefix\url{https://opg.optica.org/josab/abstract.cfm?URI=josab-35-11-2705}

\bibitem{rajnandan2023}
R.C. Das, D.~Shylla, A.~Bera, K.~Pandey, Narrow-line cooling of $^{87}$rb using
  5s1/2$\rightarrow$ 6p3/2 open transition at 420 nm.
\newblock Journal of Physics B: Atomic, Molecular and Optical Physics
  \textbf{56}(2), 025301 (2023).
\newblock \doi{10.1088/1361-6455/acabf0}.
\newblock \urlprefix\url{https://doi.org/10.1088/1361-6455/acabf0}

\bibitem{Ding2022}
R.~Ding, A.~Orozco, J.~Lee, N.~Claussen, Narrow-linewidth laser cooling for
  rapid production of low-temperature atoms for high data-rate quantum sensing.
\newblock Tech. rep., Sandia National Lab.(SNL-NM), Albuquerque, NM (United
  States) (2022)

\bibitem{rajnandan2024continuous}
R.C. Das, S.~Khan, T.~R, K.~Pandey.
\newblock Continuous loading of magneto-optical trap of rb at narrow transition
  (2024)

\bibitem{Morsch2018}
O.~Morsch, I.~Lesanovsky, Dissipative many-body physics of cold rydberg atoms.
\newblock La Rivista del Nuovo Cimento \textbf{41}(7), 383--414 (2018).
\newblock \doi{10.1393/ncr/i2018-10149-7}.
\newblock \urlprefix\url{https://doi.org/10.1393/ncr/i2018-10149-7}

\bibitem{Faoro2016PRA}
R.~Faoro, C.~Simonelli, M.~Archimi, G.~Masella, M.M. Valado, E.~Arimondo,
  R.~Mannella, D.~Ciampini, O.~Morsch, van der waals explosion of cold rydberg
  clusters.
\newblock Phys. Rev. A \textbf{93}, 030701 (2016).
\newblock \doi{10.1103/PhysRevA.93.030701}.
\newblock \urlprefix\url{https://link.aps.org/doi/10.1103/PhysRevA.93.030701}

\bibitem{Valado2016PRA}
M.M. Valado, C.~Simonelli, M.D. Hoogerland, I.~Lesanovsky, J.P. Garrahan,
  E.~Arimondo, D.~Ciampini, O.~Morsch, Experimental observation of controllable
  kinetic constraints in a cold atomic gas.
\newblock Phys. Rev. A \textbf{93}, 040701 (2016).
\newblock \doi{10.1103/PhysRevA.93.040701}.
\newblock \urlprefix\url{https://link.aps.org/doi/10.1103/PhysRevA.93.040701}

\bibitem{Levine2018PRL}
H.~Levine, A.~Keesling, A.~Omran, H.~Bernien, S.~Schwartz, A.S. Zibrov,
  M.~Endres, M.~Greiner, V.~Vuleti\ifmmode~\acute{c}\else \'{c}\fi{}, M.D.
  Lukin, High-fidelity control and entanglement of rydberg-atom qubits.
\newblock Phys. Rev. Lett. \textbf{121}, 123603 (2018).
\newblock \doi{10.1103/PhysRevLett.121.123603}.
\newblock
  \urlprefix\url{https://link.aps.org/doi/10.1103/PhysRevLett.121.123603}

\bibitem{Ebadi2021}
S.~Ebadi, T.T. Wang, H.~Levine, A.~Keesling, G.~Semeghini, A.~Omran,
  D.~Bluvstein, R.~Samajdar, H.~Pichler, W.W. Ho, S.~Choi, S.~Sachdev,
  M.~Greiner, V.~Vuleti{\'{c}}, M.D. Lukin, Quantum phases of matter on a
  256-atom programmable quantum simulator.
\newblock Nature \textbf{595}(7866), 227--232 (2021).
\newblock \doi{10.1038/s41586-021-03582-4}.
\newblock \urlprefix\url{https://doi.org/10.1038/s41586-021-03582-4}

\bibitem{Bluvstein2022}
D.~Bluvstein, H.~Levine, G.~Semeghini, T.T. Wang, S.~Ebadi, M.~Kalinowski,
  A.~Keesling, N.~Maskara, H.~Pichler, M.~Greiner, V.~Vuleti{\'{c}}, M.D.
  Lukin, A quantum processor based on coherent transport of entangled atom
  arrays.
\newblock Nature \textbf{604}(7906), 451--456 (2022).
\newblock \doi{10.1038/s41586-022-04592-6}.
\newblock \urlprefix\url{https://doi.org/10.1038/s41586-022-04592-6}

\bibitem{Salvi2023PRL}
L.~Salvi, L.~Cacciapuoti, G.M. Tino, G.~Rosi, Atom interferometry with rb blue
  transitions.
\newblock Phys. Rev. Lett. \textbf{131}, 103401 (2023).
\newblock \doi{10.1103/PhysRevLett.131.103401}.
\newblock
  \urlprefix\url{https://link.aps.org/doi/10.1103/PhysRevLett.131.103401}

\bibitem{Zhang2017}
S.~Zhang, X.~Zhang, J.~Cui, Z.~Jiang, H.~Shang, C.~Zhu, P.~Chang, L.~Zhang,
  J.~Tu, J.~Chen, {Compact Rb optical frequency standard with 10-15 stability}.
\newblock Review of Scientific Instruments \textbf{88}(10), 103106 (2017).
\newblock \doi{10.1063/1.5006962}.
\newblock \urlprefix\url{https://doi.org/10.1063/1.5006962}

\bibitem{Ling2014OL}
L.~Ling, G.~Bi, Isotope 87rb faraday anomalous dispersion optical filter at 420
  nm.
\newblock Opt. Lett. \textbf{39}(11), 3324--3327 (2014).
\newblock \doi{10.1364/OL.39.003324}.
\newblock
  \urlprefix\url{https://opg.optica.org/ol/abstract.cfm?URI=ol-39-11-3324}

\bibitem{Guan2023IEEE}
X.~Guan, W.~Zhuang, T.~Shi, J.~Miao, J.~Zhang, J.~Chen, B.~Luo, 420-nm faraday
  optical filter with 2.7-mhz ultranarrow bandwidth based on laser cooled 87rb
  atoms.
\newblock IEEE Photonics Technology Letters \textbf{35}(12), 672--675 (2023).
\newblock \doi{10.1109/LPT.2023.3272005}

\bibitem{Budker2015}
S.~Pustelny, L.~Busaite, M.~Auzinsh, A.~Akulshin, N.~Leefer, D.~Budker,
  Nonlinear magneto-optical rotation in rubidium vapor excited with blue light.
\newblock Phys. Rev. A \textbf{92}, 053410 (2015).
\newblock \doi{10.1103/PhysRevA.92.053410}.
\newblock \urlprefix\url{https://link.aps.org/doi/10.1103/PhysRevA.92.053410}

\bibitem{Boon1998}
J.R. Boon, E.~Zekou, D.J. Fulton, M.H. Dunn, Experimental observation of a
  coherently induced transparency on a blue probe in a doppler-broadened
  mismatched v-type system.
\newblock Phys. Rev. A \textbf{57}, 1323--1328 (1998).
\newblock \doi{10.1103/PhysRevA.57.1323}.
\newblock \urlprefix\url{https://link.aps.org/doi/10.1103/PhysRevA.57.1323}

\bibitem{Ogaro2020JphyB}
E.O. Nyakang’o, D.~Shylla, V.~Natarajan, K.~Pandey, Hyperfine measurement of
  the 6p1/2 state in 87rb using double resonance on blue and ir transition.
\newblock Journal of Physics B: Atomic, Molecular and Optical Physics
  \textbf{53}(9), 095001 (2020).
\newblock \doi{10.1088/1361-6455/ab7670}.
\newblock \urlprefix\url{https://dx.doi.org/10.1088/1361-6455/ab7670}

\bibitem{Ogaro2021PRA}
E.O. Nyakang'o, K.~Pandey, Resolving closely spaced levels for doppler
  mismatched double resonance.
\newblock Phys. Rev. A \textbf{103}, 013107 (2021).
\newblock \doi{10.1103/PhysRevA.103.013107}.
\newblock \urlprefix\url{https://link.aps.org/doi/10.1103/PhysRevA.103.013107}

\bibitem{Shylla2022EPJD}
D.~Shylla, E.O. Nyakang'o, R.C. Das, K.~Pandey, Effect of detuning on
  velocity-induced population oscillation.
\newblock The European Physical Journal D \textbf{76}(7), 125 (2022).
\newblock \doi{10.1140/epjd/s10053-022-00431-5}.
\newblock \urlprefix\url{https://doi.org/10.1140/epjd/s10053-022-00431-5}

\bibitem{das2006precise}
D.~Das, V.~Natarajan, Precise measurement of hyperfine structure in the 6p1/2
  state of 133cs.
\newblock Journal of Physics B: Atomic, Molecular and Optical Physics
  \textbf{39}(8), 2013 (2006)

\bibitem{banerjee2003saturated}
A.~Banerjee, V.~Natarajan, Saturated-absorption spectroscopy: eliminating
  crossover resonances by use of copropagating beams.
\newblock Optics letters \textbf{28}(20), 1912--1914 (2003)

\bibitem{das2006Epjd}
D.~Das, V.~Natarajan, Precise measurement of hyperfine structure in the 5 2 p
  1/2 state of rb.
\newblock The European Physical Journal D-Atomic, Molecular, Optical and Plasma
  Physics \textbf{37}, 313--317 (2006)

\bibitem{Gerhardt1978}
H.~Gerhardt, E.~Matthias, F.~Schneider, A.~Timmermann, Isotope shifts and
  hyperfine structure of the 6s-7p transitions in the cesium isotopes 133, 135,
  and 137.
\newblock Zeitschrift f{\"u}r Physik A Atoms and Nuclei \textbf{288}(4),
  327--333 (1978).
\newblock \doi{10.1007/BF01417714}.
\newblock \urlprefix\url{https://doi.org/10.1007/BF01417714}

\bibitem{Wang2013Cs}
and, , , , and, Cs 455 nm nonlinear spectroscopy with ultra-narrow linewidth.
\newblock Chinese Physics Letters \textbf{30}(6), 060601 (2013).
\newblock \doi{10.1088/0256-307X/30/6/060601}.
\newblock \urlprefix\url{https://dx.doi.org/10.1088/0256-307X/30/6/060601}

\bibitem{Miao2022Cs}
J.~Miao, T.~Shi, J.~Zhang, J.~Chen, Compact 459-nm cs cell optical frequency
  standard with
  $2.1\ifmmode\times\else\texttimes\fi{}{10}^{\ensuremath{-}13}/\sqrt{\ensuremath{\tau}}$
  short-term stability.
\newblock Phys. Rev. Appl. \textbf{18}, 024034 (2022).
\newblock \doi{10.1103/PhysRevApplied.18.024034}.
\newblock
  \urlprefix\url{https://link.aps.org/doi/10.1103/PhysRevApplied.18.024034}

\bibitem{Zhang2016Cs1}
X.~Zhang, Z.~Jiang, Z.~Tao, H.~Shang, C.~Zhang, J.~Chen, \emph{Research on Cs
  active Faraday optical frequency standard with 459 nm laser pumping}, in
  \emph{2016 IEEE International Frequency Control Symposium (IFCS)} (2016), pp.
  1--3.
\newblock \doi{10.1109/FCS.2016.7546767}

\bibitem{Zhang2016Cs2}
S.~Zhang, X.~Zhang, Z.~Jiang, D.~Pan, X.~Peng, H.~Chen, J.~Chen, H.~Guo,
  \emph{A compact optical clock scheme based on caesium atomic beam}, in
  \emph{2016 IEEE International Frequency Control Symposium (IFCS)} (2016), pp.
  1--4.
\newblock \doi{10.1109/FCS.2016.7546771}

\bibitem{Glaser2020}
C.~Glaser, F.~Karlewski, J.~Kluge, J.~Grimmel, M.~Kaiser, A.~G\"unther,
  H.~Hattermann, M.~Krutzik, J.~Fort\'agh, Absolute frequency measurement of
  rubidium $5s\text{\ensuremath{-}}6p$ transitions.
\newblock Phys. Rev. A \textbf{102}, 012804 (2020).
\newblock \doi{10.1103/PhysRevA.102.012804}.
\newblock \urlprefix\url{https://link.aps.org/doi/10.1103/PhysRevA.102.012804}

\bibitem{Hemmerich1990OL}
A.~Hemmerich, D.H. McIntyre, C.~Zimmermann, T.W. H\"{a}nsen, Second-harmonic
  generation and optical stabilization of a diode laser in an external ring
  resonator.
\newblock Opt. Lett. \textbf{15}(7), 372--374 (1990).
\newblock \doi{10.1364/OL.15.000372}.
\newblock
  \urlprefix\url{https://opg.optica.org/ol/abstract.cfm?URI=ol-15-7-372}

\bibitem{hayasaka2002}
K.~Hayasaka, Frequency stabilization of an extended-cavity violet diode laser
  by resonant optical feedback.
\newblock Optics Communications \textbf{206}(4), 401--409 (2002).
\newblock \doi{https://doi.org/10.1016/S0030-4018(02)01446-3}.
\newblock
  \urlprefix\url{https://www.sciencedirect.com/science/article/pii/S0030401802014463}

\bibitem{Chang2019}
P.~Chang, S.~Zhang, H.~Shang, J.~Chen, Stabilizing diode laser to 1 hz-level
  allan deviation with atomic spectroscopy for rb four-level active optical
  frequency standard.
\newblock Applied Physics B \textbf{125}(11), 196 (2019).
\newblock \doi{10.1007/s00340-019-7313-x}.
\newblock \urlprefix\url{https://doi.org/10.1007/s00340-019-7313-x}

\end{thebibliography}
%%\input bibDirectSpectro.bbl

\end{document}